\documentclass[aip,jap,amsmath,amssymb,preprint]{revtex4-1}

\usepackage{graphicx}
\usepackage{dcolumn}
\usepackage{bm}

\begin{document}

\preprint{AIP/123-QED}

\title[]{The photoionization cross section of a hydrogenic impurity in a multi-layered spherical quantum dot}

\author{Mehmet \c{S}ahin}
\email{mehmet.sahin@agu.edu.tr, sahinm@selcuk.edu.tr}
\affiliation{Department of Material Science and Nanotechnology Engineering, Abdullah G\"{u}l University, Kayseri, Turkey}
\affiliation{Department of Physics, Faculty of Science, Sel\c{c}uk University, Campus 42075 Konya, Turkey}
\author{Firdes Tek}
\affiliation{Department of Physics, Faculty of Science, Sel\c{c}uk University, Campus 42075 Konya, Turkey}

\author{Ahmet Erdin\c{c}}
\affiliation{Department of Physics, Faculty of Science, Erciyes University, 38039 Kayseri, Turkey}


\begin{abstract}
In this study, we have investigated the photoionization cross section of an on-center hydrogenic impurity in a multi-layered spherical quantum dot. The electronic energy levels and their wave functions have been determined fully numerically by shooting method. Also, we have calculated the binding energy of the impurity by using these energy values. The photoionization cross section has also been computed as a function of the layer thickness and normalized photon energies. We have discussed in detail the possible physical reasons behind the changes in the binding energies and photoionization cross section. It is observed that both the binding energies and the photoionization cross sections depend strongly on the layer thickness and photon energies.
%
\end{abstract}

\maketitle

\section{Introduction}

An ionization process is removal of one or more electrons from a quantum system, such as atom, molecule, quantum heterostructure etc., in any manners. If the ionization process is induced by a photon, this process is called photoionization. The photoionization of atomic systems provides the opportunity to investigate the dynamic interplay among many body electron-electron correlations and relativistic effects.\cite{kje} It is the simplest process giving detailed information on the atomic and molecular structure.\cite{raf} Numerous experimental and theoretical studies on photoionization cross section of different atoms have been published by many authors.\cite{sale,kje1,sch} Photoionization cross sections are of great significance due to its numerous applications in space research, astrophysics, radiation protection, laser designing, controlled thermonuclear research and different types of laboratory plasma.\cite{raf,kje}

On the other hand, because quantum dot (QD) heterostructures exhibit atomic properties such as energy levels, density of states etc., they are called as artificial atoms.\cite{sah} The investigation of photoionization cross section of a hydrogenic impurity in a QD is very important especially for different optoelectronic device applications such as quantum dot infrared photodetectors. Although there are no experimental studies, the photoionization cross section of a hydrogenic impurity in low-dimensional quantum structures has been studied theoretically by many authors.\cite{sah,bar,ham1,sai,kaw,ila,sal,ham2,cor} The cross section strongly depends on the impurity binding energy and its wave function.\cite{ham1} The photoionization cross section is still studied extensively under different physical parameters (i.e electric field, magnetic field, hydrostatic pressure) in low-dimensional quantum nano structures. Recently, \c{S}ahin\cite{sah} has investigated the photoionization cross section of both $D^0$ and $D^-$ impurities in a QD for an on-center impurity as a function of the dot radii and the normalized photon energies. He demonstrated that the cross section drastically depends on the dot size, the photon energy and the number of electrons in the QD. Barseghyan et al.\cite{bar} studied the behavior of the binding energy and the photoionization cross-section of a donor impurity in a cylindrical-shape GaAs/AlGaAs quantum dot, under the effects of hydrostatic pressure and applied electric and magnetic fields in-growth direction. They showed that when the hydrostatic pressure is increased, the photoionization cross section grows. Ham and Spector\cite{ham1} analyzed the dependence of the photoionization cross section on the energy and polarization of the photons in a spherical QD as a function of dot radius and impurity locations for both infinite and finite potential barriers. They found that the cross section is independent of the polarization of the photons for an on-center impurity, while it depends on the polarization of the photon field for an off-center one. Sajeev and Moiseyev\cite{saj} investigated the autoionization and photoionization in two-electron spherical quantum dots. Baldea and Koppel\cite{bal} showed that the photoionization is a valuable method, which can be used along with transport measurements, to study single electron transistors. Most recently, Lin and Ho\cite{lin} have studied the photoionization cross section of a hydrogen impurity in spherical quantum dots which are modelled by finite oscillator and Gaussian potentials. All of these studies have been reported for a single core/shell QD.

The production of the spherical multi-layered quantum dot (MLQD) has become possible using wet chemical techniques.\cite{dor,mew} The electronic properties of multilayered spherical quantum dots under various physical effects have been investigated by some authors.\cite{boz0,boz1,akt}  These studies are related to electronic energy level and impurity binding energy calculations. It is seen from these studies that the energy levels and impurity binding energies are very sensitive to the layer thickness. Until now, to the best of our knowledge, although the electronic structure has been studied by many authors, the photoionization cross section properties have not been investigated in a MLQD. The investigation of photoionization cross section in MLQD may give rise to an opportunity for producing a new generation of quantum devices such as laser source, photodetectors, single electron transistor etc. For example, a single electron transistor can work as depending on the photoionization cross section.\cite{bal} The controllability of the shell thicknesses may provide different advantages in the production of these kind of devices working based on the photoionization cross section.

The main aim of this work is to calculate the photoionization cross section of a hydrogenic impurity in a spherical multi-layered semiconductor quantum dot, and to investigate the dependence of the photoionization on the layer thickness. For this purpose, the ground and first excited energy levels and corresponding wave functions are calculated by solving the Schr\"{o}dinger equation with shooting method for an electron confined in a MLQD. The variations of the impurity binding energies and the photoionization cross section are examined as a function of shell thicknesses and photon energies.

The rest of this paper is organized as follows: In the next section, the model and calculations are presented. The calculation results and discussion of their probable reasons are given in section III. A brief conclusion is presented in the last section.

\section{Model and Calculation}

\begin{figure}
\includegraphics[width=3.4in]{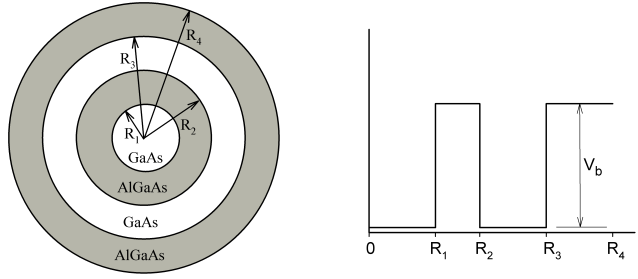}
\caption{\label{fig:1} Schematic representation and potential profile of a core/shell/well/shell QD.}
\end{figure}

In this study, a spherically symmetric core/shell/well/shell quantum dot structure is considered. As seen from Fig. \ref{fig:1}, GaAs core is a sphere of radius $R_{1}$. The well region is that another GaAs spherical shell of inner radius $R_{2}$ and outer radius $R_{3}$ is concentric with the core region. The gap between the core and well regions is filled with AlGaAs as a shell (barrier) zone which isolates the core and well regions and also allows electron to tunnel between these regions. The shell thickness is $T_s=R_{2}-R_{1}$ and the well region thickness is $T_w=R_{3}-R_{2}$. All structure is embedded in an AlGaAs matrix material.


In the effective mass approximation, for a spherically symmetric quantum dot, the single-particle Schr\"{o}dinger equation is given as

\begin{widetext}
\begin{equation}
\label{eq1}
\left[ {- \frac{{\hbar ^2}}{2}\vec \nabla_r \left( {\frac{1}{{m^* \left(r \right)}}\vec \nabla _r } \right) -
\frac{{Ze^2 }}{\kappa (r){\left| {\vec r - \vec r_i } \right|}} + \frac{{\ell \left( {\ell  + 1} \right)\hbar^2 }}{{2m^* \left( r \right)r^2 }}+ V(r)} \right]R_{n,\ell } \left( r \right) = \varepsilon _{n,\ell} R_{n,\ell} \left( r \right),
\end{equation}
\end{widetext}

\noindent
where $\hbar$ is reduced Planck constant, $m^{\ast}(r)$ is the position-dependent electron's effective mass, $Z$ is charge of the impurity, $\kappa (r)$ is the position dependent dielectric constant, $\vec r_i$ is the position of the impurity, $\ell$ is the angular momentum quantum number, $V(r)$ is the finite confining potential, $\varepsilon_{n,\ell}$ is the electron energy eigenvalue, and $R_{n,\ell}(r)$ is the radial wave functions of the electron. It should be noted that, in the case without the impurity, the charge of the impurity is $Z=0$ and in the existence of the impurity, it is $Z=1$. In the calculations, we have taken into consideration an on-center impurity and hence the $\vec r_i=0$. The mathematical expression of the confining potential is

\begin{equation}
 \label{eq2}
 V(r)=\left\{{\begin{array}{l}
 0,\,\,\,\,\,\,\,\,\,\,\,\,\,r \leq R_1 \,\,\, $and$ \,\,\, R_2\leq r \leq R_3\\\\
 V_b, \,\,\,\,\,\,\,\,\,\,R_1<r<R_2 \,\,\, $and$ \,\,\, r>R_3 \\
 \end{array}} \right..
\end{equation}

In order to determine the single particle energy levels and corresponding wave functions, Eq.(\ref{eq1}) is full numerically solved by using the shooting method. As is well known, this technique converts an eigenvalue problem to an initial value problem. For this purpose, Hamiltonian operator is discretized on a uniform radial mesh in 1D using the finite differences, then Eq.(\ref{eq1}) can be reduced to an initial value equation. Here, $\Delta r$ width between two mesh points is chosen as 0.005. All details of this method can be found in Ref.\onlinecite{har}.

\subsection{Theory of photoionization cross-section}

The photoionization process is an optical transition that takes place from the impurity ground state as the initial state to the conduction subbands (continuum), which starts above the confining potential of the dot. The sufficient energy is required in order for the transition to occur\cite{sah,sal}. The photoionization cross section can be defined as the ionization probability of the electrons from the bound state under an external optical excitation. It is strongly dependent on the confinement potential and the photon energy. The excitation energy depends on the photoionization cross section associated with an impurity and, as in the bulk case, it can be calculated in frame of the Fermi's golden rule and the dipole approximation as following\cite{lax}

\begin{widetext}
\begin{equation}
\label{eq3}
\sigma \left( {\hbar \omega } \right) = \left[ {\left(
{\frac{{F_{eff} }} {{F_0 }}} \right)^2 \frac{{n_r }} {\kappa }}
\right]\frac{{4\pi ^2 }} {3}\beta _{FS} \hbar \omega \sum\limits_f
{\left| {\left\langle {\psi _i \left| \textbf{r} \right|\psi _f }
\right\rangle } \right|} ^2 \delta \left( {E_f  - E_i  - \hbar
\omega } \right),
\end{equation}
\end{widetext}

\noindent
where $n_{r}$ is the refractive index of the semiconductor, $\kappa$ is the dielectric constant of the medium, $\beta_{FS}=e^{2}$/ $\hbar c$ is the fine structure constant, and $\hbar\omega$ is the photon energy. $F_{eff}/F_{0}$ is the ratio of the effective electric field $F_{eff}$ of the incoming photon and average field $F_{0}$ in the medium\cite{rid}. $<\psi_{i}|\textbf{r}|\psi_{f}>$ is the matrix element between the initial and final states of the dipole moment of the impurity. $\psi_{f}$ and $\psi_{i}$ are the wave functions of the final and initial states, and $E_{f}$ and $E_{i}$ are the corresponding energy eigenvalues of these states, respectively. These eigenvalues and eigenvectors are determined by the shooting method.

In a spherical QD, the selection rules ($\Delta\ell=\pm1$) determine the final state of the electron after the impurity photoionization. The 1$s$ impurity state is taken as ground level for $D^{0}$ and the 1$p$ level of one electron system without the impurity is taken as the final state for $D^{0}$ as similar with previous studies.\cite{sah,ham1,cor,sai,kaw,ila}

In order to calculate the numerical values of the photoionization cross-section given by Eq.(\ref{eq4}), $F_{eff}/F_{0}$ is taken as approximately unity because the calculation of this quantity is very difficult and it has no effect on the photoionization cross-section shape\cite{ila,sai,kaw,sal}. Also, the initial and final state wave functions ($\psi_{i}$ and $\psi_{f}$) are determined by the multiplication of the radial wave function with the spherical harmonics i.e. $R_{n,\ell}(r)Y_{\ell,m}(\theta,\varphi)$. Here, $Y_{\ell,m}(\theta,\varphi)$ is the spherical harmonics. In addition, the $\delta$-function in Eq.(\ref{eq4}) is replaced by a narrow Lorentzian by means of,

\begin{equation}
\label{eq4}
\delta \left( {E_f  - E_i  - \hbar \omega } \right) =
\frac{{\hbar \Gamma }} {{\pi \left[ {\left( {\hbar \omega  - \left(
{E_f  - E_i } \right)} \right)^2  + \left( {\hbar \Gamma } \right)^2
} \right]}}.
\end{equation}

\noindent
Here, $\Gamma$ is the hydrogenic impurity linewidth and taken as 0.1$R_y^*$.

\section{Results and Discussion}

The atomic units have been used throughout the calculations, where $\hbar=m=e=1$. Effective Bohr radius is $a_0^*\simeq100$ {\AA} and effective Rydberg energy is $R_y^*\simeq5.25$ meV. The confinement potential has been taken as $V_b=228$ meV. The material parameters have been taken as $m_{GaAs}=0.067\ m_0$, $m_{AlGaAs}=0.088\ m_{0}$, $\kappa_{GaAs}=13.18$, $\kappa_{AlGaAs}=12.8$. Also the effective masses of electrons inside GaAs and AlGaAs are $m_1^*$ and $m_2^*$, and the dielectric constants are $\kappa_1$ and $\kappa_2$, respectively. The position-dependent effective mass and the dielectric constant may be defined as follows\cite{buc}

\begin{eqnarray}
\label{eq5}
\nonumber
m^\ast(r) = \left\{ {\begin{array}{l}
 1,\,\,\,\,\,\,\,r \leq R_1 \,\,\, $and$ \,\,\, R_2\leq r \leq R_3 \\\\
 \frac{m_2^\ast }{m_1^\ast },\,\, R_1<r<R_2 \,\,\, $and$ \,\,\, r>R_3 \\
 \end{array}} \right. \\ \nonumber
\\
\kappa(r) = \left\{ {\begin{array}{l}
 1,\,\,\,\,\,r \leq R_1 \,\,\, $and$ \,\,\, R_2\leq r \leq R_3 \\\\
 \frac{\kappa _2 }{\kappa _1 },\,\,R_1<r<R_2 \,\,\, $and$ \,\,\, r>R_3 \\
 \end{array}} \right. .
\end{eqnarray}

The binding energy of a neutral donor impurity ($D^0$) in a QD is given as \cite{sah}

\begin{equation}
\label{eq6}
E_b(D^0)=E_{10}-E_{10}(D^0).
\end{equation}

\noindent
Here, $E_{10}$ is ground state energy of the single electron QD without the impurity and $E_{10}(D^0)$ is that of the neutral donor impurity.

\begin{figure}
\includegraphics[width=3.0in]{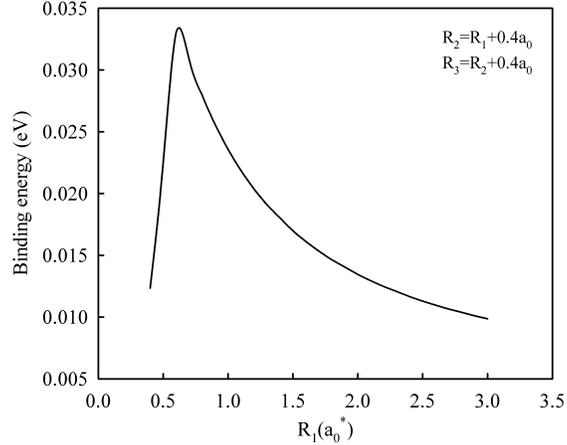}
\caption{\label{fig:2} Impurity binding energy of a multi-layered QD as a function of core radius, $R_1$, for constant $T_s$ and $T_w$.}
\end{figure}

Figure \ref{fig:2} shows the variation of the binding energy of the $D^0$ impurity as a function of the core radius of the MLQD, $R_1$. Here, the shell thickness is $T_s=0.4\ a_0^*$ and the well width is $T_w=0.4\ a_0^*$. As seen from the figure, in small core radius, the binding energy increases rapidly with increasing $R_1$. It reaches a maximum value at about $R_1=0.6\ a_0^*$ and decreases smoothly with further increasing of the $R_1$ and goes to a constant value. This behavior is rather similar to the binding energy of a neutral hydrogenic donor impurity in a single QD.\cite{sah,por}

\begin{figure}
\includegraphics[width=3.0in]{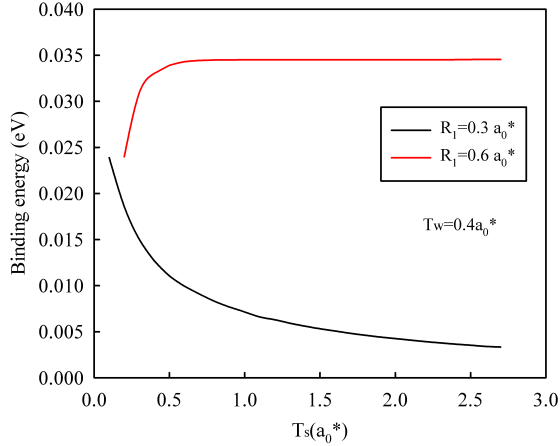}
\caption{\label{fig:3} (Color online) Impurity binding energy of a core/shell/well/shell QD as a function of shell thickness for constant well width and two different core radii.}
\end{figure}

\begin{figure}
\includegraphics[width=3.4in]{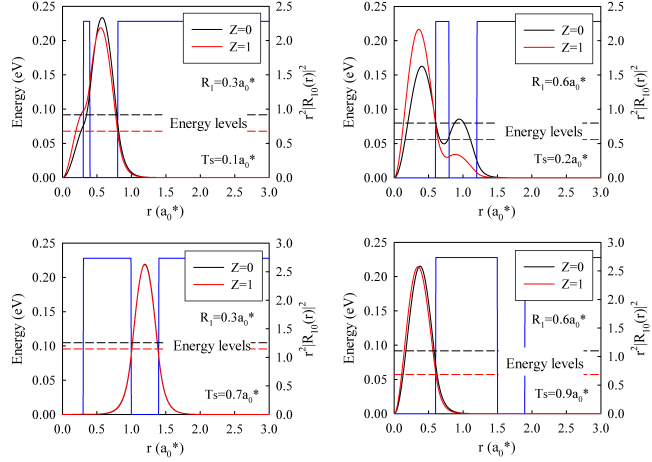}
\caption{\label{fig:4} (Color online) Ground state energies and probability densities at same $T_w=0.4\ a_0^*$ for $Z=0$ and $Z=1$ cases and different layer thickness. The other layer parameters are given on the figure.}
\end{figure}

Figure \ref{fig:3} shows the binding energy of the impurity as a function of the shell thickness for $T_w=0.4\ a_0^*$, and two different core radii, $R_1=0.3$ and $0.6\ a_0^*$. As seen from the figure, the behaviors of the binding energies are completely different from the each other. When the core radius is $R_1=0.3\ a_0^*$, the binding energy decreases with increasing shell thickness. On the other hand, when the core radius is $R_1=0.6\ a_0^*$, the binding energy increases firstly and remains at a constant value for almost $T_s>0.6\ a_0^*$. These changes are explained as follows: In the case of small core radius, the electron is confined in the well region for both cases without ($Z=0$) and with ($Z=1$) the impurity because the core region is rather small. However, in the presence of the impurity, the electron feels the effect of the impurity at smaller shell thickness and hence the energy level is pulled down because of the attractive coulomb potential of the impurity. As a result of this, the binding energy becomes larger at small shell thickness. In contrast, while the shell  thickness increases, the attractive effect of the impurity on the electron decreases so that, the electron is less affected from the impurity. These behaviors are clearly seen from the left panel of Fig.\ref{fig:4}. Accordingly, the binding energy decreases with increasing $T_s$. By further increase in the $T_s$, the electron is confined completely in the well region for both $Z=0$ and $Z=1$ cases, and it is insubstantially affected from the impurity. In the case of large core radius, while the electron is confined in both the core and well regions for $Z=0$ case, it is almost confined in the core region for $Z=1$ case. As a result, the energy difference is relatively small. When the shell thickness, $T_s$, is larger, the electron is completely confined in the core region for both $Z=0$ and $Z=1$ cases and the effect of the well region vanishes as seen from the right panel of Fig.\ref{fig:4}. Hence, the attractive coulomb effect of the impurity becomes more significant which results in the larger binding energy as seen in Fig. \ref{fig:3}.

\begin{figure}
\includegraphics[width=3.0in]{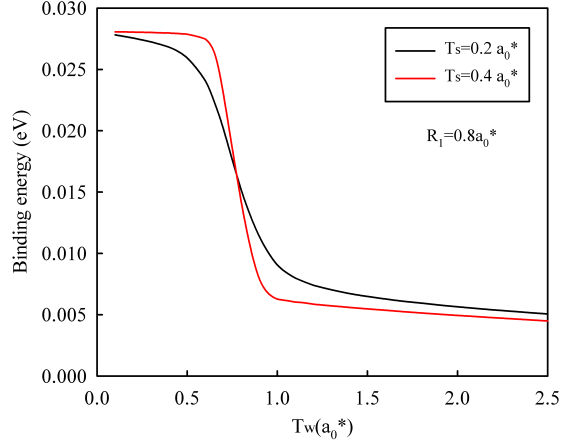}
\caption{\label{fig:5} (Color online) Variation of the impurity binding energy as a function of the well widths, $T_w$, for $R_1=0.8\ a_0^*$, and $T_s=0.2$ and $0.4\ a_0^*$.}
\end{figure}

\begin{figure}
\includegraphics[width=3.4in]{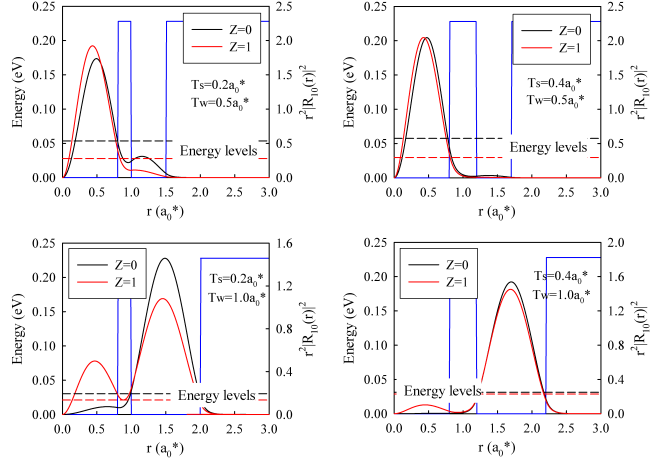}
\caption{\label{fig:6} (Color online) Ground state energies and probability densities at same $R_1=0.8\ a_0^*$ for $Z=0$ and $Z=1$ cases and different layer thickness.}
\end{figure}

We expect that the impurity binding energy will be affected drastically from the width of well region. In Fig. \ref{fig:5}, the impurity binding energies are plotted as a function of the well width, $T_w$, for two different shell thickness at $R_1=0.8\ a_0^*$. When we look at the figure, we see similar treatments in the binding energies for both shell thicknesses. Nevertheless, we observe that the variation of the binding energy is smoother for $T_s=0.2\ a_0^*$ as compared with case of $T_s=0.4\ a_0^*$. The physical reasons of these treatments can be explained as follows: For the first value of the shell thickness, in spite of the fact that the electron is confined inside the core region for both $Z=0$ and $Z=1$ cases, the well region influences the energy levels even for small values of the well width. Moreover, the energy difference between $Z=0$ and $Z=1$ cases is larger, so that the binding energy increases. When the well region becomes much larger, the electron tunnels easily to the well region and the energy levels approaches to each other. This situation is shown in left panel of Fig.\ref{fig:6}. On the other hand, when the shell thickness is $T_s=0.4\ a_0^*$, the tunneling is not so easy and the electron is confined almost completely in the core region in comparison with the $T_s=0.2\ a_0^*$ case as seen in right panel of Fig.\ref{fig:6}. As opposed to the $T_s=0.2\ a_0^*$ case, when the well region is much larger, the electron tunnels completely to the well region and as depending on this, the decreasing in the binding energy becomes sharper with increasing shell thickness as observed in Fig. \ref{fig:5}.

The photoionization cross section of a donor impurity is an important phenomena especially for device applications of quantum heterostructures. If the photon energy is equal to the energy difference between the initial and the final states (i.e. resonant energy, $\hbar \omega=E_f-E_i$), all terms of Eq.\ref{eq3}, except that photon energy and dipole matrix element, contribute as a constant to the photoionization cross section. In this resonant photon energy case, the Lorentz function becomes also another constant as $1/\pi\hbar\Gamma$. On the other hand, the energy difference between the initial and the final states and the overlapping of their wave functions (hence the dipole matrix element) play an effective role on the shape of the photoionization cross section.\cite{sah}

\begin{figure}
\includegraphics[width=3.0in]{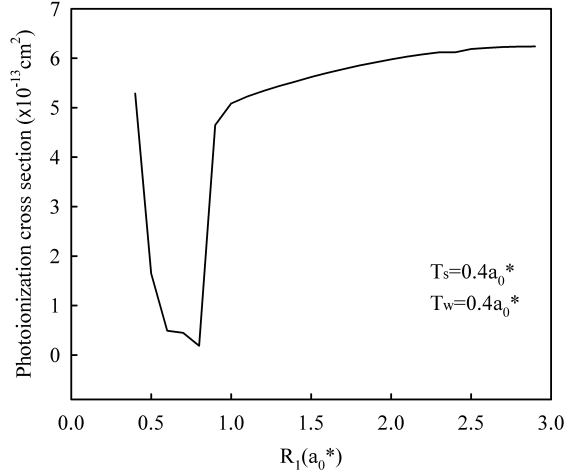}
\caption{\label{fig:7} The photoionization cross section of the hydrogenic donor impurity vs. core radius at constant $T_s$ and $T_w$.}
\end{figure}

Fig. \ref{fig:7} shows the variation of the peak values of the photoionization cross section with the core radius, $R_1$, for constant shell and well widths, i.e. $T_s=0.4\ a_0^*$ and $T_w=0.4\ a_0^*$. These peak values correspond to the resonant photon energies. As seen from the figure, while the photoionization has very large value at $R_1=0.4\ a_0^*$, it decreases too sharp with increasing $R_1$ and reaches the its smallest value at $R_1=0.8\ a_0^*$. When the core radius is $R_1=0.9\ a_0^*$, it increases instantly and remains stable with further increasing of the $R_1$. The physical reason of this treatment is as follows: at very small $R_1$ values, the wave functions of both ground and excited states are localized in the well region as similar to Fig. \ref{fig:4} and hence the overlapping of the wave functions becomes larger. As a result of this, the photoionization cross section is also larger.In this situation, the photoionization takes place in the well region and this may be important for a photodetector application which is sensitive to higher photon energies (blue shift) relatively, because the energy levels become larger, in smaller core radii.The increase in the core radius causes a decrease in the overlapping of the wave functions because the impurity wave function is confined in the core region while the excited states wave function is still confined in the well region. Therefore, the photoionization cross section has small values. With further increasing of the core radius, both ground and excited states wave functions are localized in the core region and hence the overlapping of the wave functions becomes large and as depending on this, the cross section is also to be maximum. In this case, the photoionization occurs in the core region. This situation may be important to produce a photodetector which is sensitive to lower energies (red shift).

\begin{figure}
\includegraphics[width=3.0in]{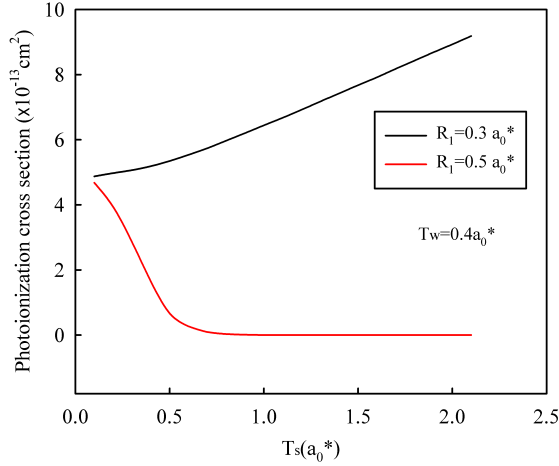}
\caption{\label{fig:8} (Color online) The variation of the photoionization cross section of the impurity depending on shell thickness for two different core radii and constant $T_w$.}
\end{figure}

Figure \ref{fig:8} shows the variation of the photoionization cross section as a function of the shell thickness, $T_s$, for two different values of the core radius, $R_1$, and the same well thickness, $T_w=0.4\ a_0^*$. When we look at the figure, we see that the cross section increases with increasing shell thickness for $R_1=0.3\ a_0^*$. On the other hand, for $R_1=0.5\ a_0^*$, in contrast with the previous case, the cross section exhibits completely different behavior and it decreases rapidly with increasing $T_s$. Similar treatments are also observed for $R_1>0.5\ a_0^*$ values. When $R_1=0.3\ a_0^*$, the wave functions of ground and excited states are confined in the well region and the finding probability in the core region vanishes almost for large values of the shell thickness. Therefore, the overlap between the wave functions becomes much larger in the well region as similar to the left panel of Fig. \ref{fig:4}. This is the reason of an increase in the photoionization cross section which happens in the well region. When $R_1=0.5\ a_0^*$, ground state wave function is confined in the core region because of the attractive coulomb potential of the impurity. At small values of $T_s$, the tunneling of the excited state wave function to the core region is much more. In this case, the overlapping of the wave functions becomes larger and hence the photoionization cross section becomes greater. When the barrier thickness increases, the overlapping of the wave functions decreases because of the localizing of excited state wave function in the well region and therefore the cross section becomes less. We conclude that the sensitivity of a device can be also tuned with the shell thickness of MLQD to a narrower or larger region.

\begin{figure}
\includegraphics[width=3.0in]{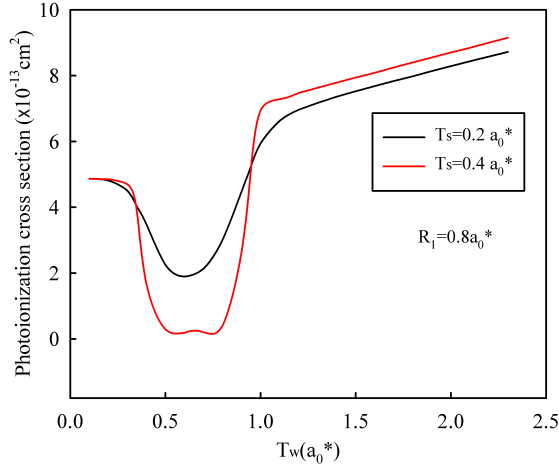}
\caption{\label{fig:9} (Color online) The variation of the photoionization cross section of the impurity as a function of well width for two different shell thickness and constant core radius.}
\end{figure}

\begin{figure}
\includegraphics[width=3.0in]{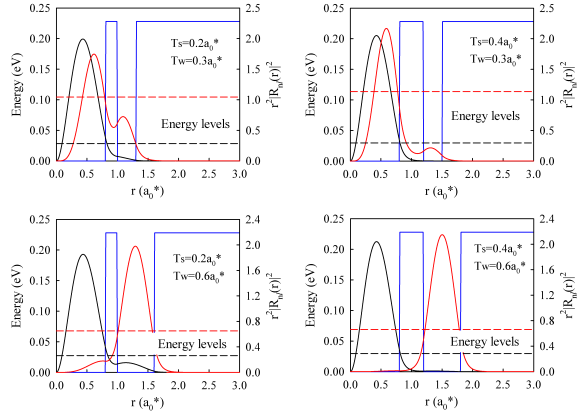}
\caption{\label{fig:10} (Color online) Localization and energy cases for ground state of Z=1 (black lines) and excited state of Z=0 (red lines). The left panel for $T_s=0.2\ a_0^*$ and the right panel for $T_s=0.4\ a_0^*$.}
\end{figure}

The variation of the photoionization cross section with the well width, $T_w$, is seen in Fig. \ref{fig:9} for $R_1=0.8\ a_0^*$ and two different shell thickness, $T_s=0.2$ and $0.4\ a_0^*$. As seen from the figure, in small well widths, the photoionization cross section remains a bit stable and then it decreases rapidly with increasing well widths and reaches the minimum values at about $T_w=0.6\ a_0^*$. The minimum value of the photoionization for $T_s=0.4\ a_0^*$ becomes smaller than that for $T_s=0.2\ a_0^*$. The cross section increases again for both cases with further increasing of the $T_w$. We can explain this behavior as follows: In small $T_w$ values, namely narrow well region case, the wave functions of ground and excited states are localized in the core region and therefore the overlapping of them becomes greater. The photoionization occurs in the core region. The excited state wave function starts to tunnel through the well region with increasing well width and it is completely localized in this region when the well width is about $T_w=0.6\ a_0^*$ as seen from Fig. \ref{fig:10}. In this case, the overlapping of the wave functions is minimum and hence, the photoionization cross section is very small. This decreasing is much more for $T_s=0.4\ a_0^*$, because the tunneling of excited state wave function to the core region is smaller as shown at bottom panel of Fig. \ref{fig:10}. By further increase of the well width, the ground state wave function also begins to tunnel to the well region and as a result of this, the overlapping of the wave functions increases again. When $T_w\geq1.0\ a_0^*$, both of the wave functions are localized in the well region and the cross section is to be higher. The photoionization cross section occurs in the well region anymore.

\begin{figure}
\includegraphics[width=3.0in]{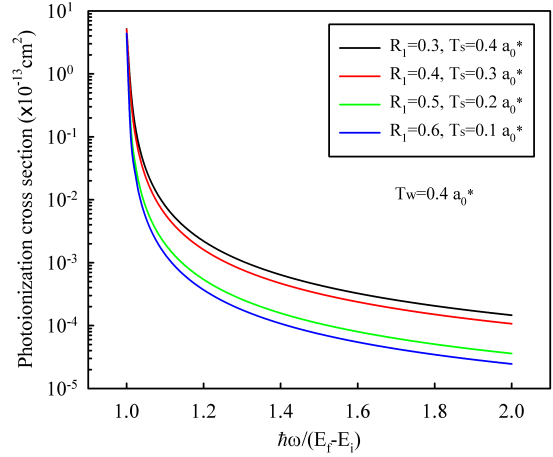}
\caption{\label{fig:11} (Color online) Photoionization cross section of the hydrogenic donor impurity as a function of normalized photon energy for different $R_1$, and $T_s$ and constant $T_w$.}
\end{figure}

The investigation of the photoionization cross section as depending on the photon energy is another important process. Figure \ref{fig:11} shows the variation of the photoionization cross section of the neutral donor impurity as a function of the normalized photon energy $\hbar\omega/(E_f-E_i)$ for different core radii, $R_1$, and shell thickness, $T_s$. As seen from the figure, the maximum photoionization corresponds to the threshold photon energy for all $R_1$ and $T_s$ values and it decreases with increasing photon energy. These general trends are quite similar to those in single QDs.\cite{sah,ila,sal,sal1} Also, the cross sections go towards lower values with increasing $R_1$ and decreasing $T_s$. When $R_1$ is small, both ground and excited state wave functions are confined in the well region which results in a strong overlap between the wave functions and therefore in a larger photoionization cross section. The increase in the core radius, which corresponds to the slimming of the shell wall, leads to a possibility of tunneling of ground state wave function through the core region due to the attractive Coulomb potential of the impurity. So this possibility reduces the value of dipole matrix element and also as depending on this, the cross section becomes smaller with increasing normalized photon energy.

\begin{figure}
\includegraphics[width=3.0in]{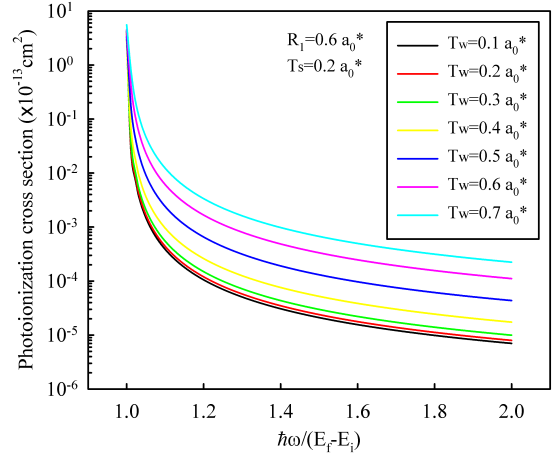}
\caption{\label{fig:12} (Color online) Photoionization cross section of the hydrogenic donor impurity as a function of normalized photon energy for different $T_w$ and constant $R_1$, and $T_s$.}
\end{figure}

In Fig. \ref{fig:12}, variation of the photoionization cross section with normalized photon energy is shown for different well widths at $R_1=0.6\ a_0^*$ and $T_s=0.2a_0^*$. As seen from the figure, the maximum photoionization cross section corresponds to the threshold energy and decreases with increasing photon energies for all well widths. When the well width increases, the photoionization cross section increases and this increase is more obvious for $T_w>0.4\ a_0^*$. This is originated from the overlapping of the wave functions in the well region. That is, the photoionization happens in the well region.

As seen from Figs. \ref{fig:11} and \ref{fig:12}, the maximum absorption occurs at the photoionization threshold and the photoionization cross section decreases monotonically from the maximum value with increasing photon energy like the true hydrogenic model. This transition requires higher photon energies than those in quantum wells and quantum wires.\cite{sal,sal1} The reason for this behavior is that the electron is completely confined in all spatial dimensions in QD structures. This interesting result, which seems to be the characteristic and the signature of the QDs, is not observed in quantum well and quantum wire structures.\cite{sah,sal,sal1}

\section*{Conclusion}

In this study, a detailed examination of the electronic structure and photoionization cross section of a donor impurity have been performed for a multi-layered spherical QD. In calculations, we have used shooting method with finite difference techniques and effective mass approximation for determining energy eigenvalues and their wave functions. First, we have calculated the donor binding energies as a function of the layer thickness. We have found that the donor binding energy strongly depends on the layer thickness. Then, we have computed photoionization cross section of the donor impurity as a function of layer thickness and normalized photon energies. We have showed that the photoionization cross section can be controlled with adjusting the core radius or layer widths. These properties can be useful for producing some device applications such as quantum dot infrared photodetectors. Although there is no experimental or theoretical reports in the literature to compare our results, general trend of our results are in a good agreement with previous studies of single quantum dots. We believe that this study will be rather useful and will contribute to the understanding of the photoionization cross section of MLQDs. We hope that this study will also stimulate both experimental and theoretical investigations of the photoionization cross section of MLQDs.

\begin{acknowledgments}
This work is a part of the M.Sc. degree thesis that was presented by F. Tek at Physics Department of Selcuk University. This study was partially supported by Selcuk University BAP office.
\end{acknowledgments}

\end{document}